\documentclass[preprint,preprintnumbers,prd,floatfix,superscriptaddress,nofootinbib] {revtex4-1}
\usepackage{epsfig}
\usepackage{amssymb}
\usepackage{amsmath}
\usepackage{dcolumn}
\usepackage{cancel}
\usepackage[colorlinks]{hyperref}
\usepackage[usenames,dvipsnames]{color}
\usepackage{epstopdf}
\hypersetup{
     breaklinks=true,
    pdfstartview={FitH},    
    colorlinks=true,       
    linkcolor=blue,          
    citecolor=red,        
    filecolor=magenta,      
    urlcolor=blue,           
    anchorcolor=green,      
    linktocpage=true
}

\def\doi{http://doi.org}

\def\d{\mathrm{d}}

\begin{document}

 \title{Deflation of the cosmological constant associated with inflation and dark energy}

 \author{Chao-Qiang Geng}
\email{geng@phys.nthu.edu.tw}
\affiliation{Chongqing University of Posts \& Telecommunications, Chongqing, 400065, 
China}
 \affiliation{National Center for Theoretical Sciences, Hsinchu,
Taiwan 300}
\affiliation{Department of Physics, National Tsing Hua University,
Hsinchu, Taiwan 300}

\author{Chung-Chi Lee}
\email{chungchi@mx.nthu.edu.tw}
 \affiliation{National Center for Theoretical Sciences, Hsinchu,
Taiwan 300}

\begin{abstract}
In order to solve the fine-tuning problem of the cosmological constant, we propose a simple model with the vacuum energy 
non-minimally coupled to the inflaton field. In this model, the vacuum energy decays to the inflaton during pre-inflation and 
inflation eras, so that the cosmological constant effectively deflates from the Planck mass scale to a much smaller one after 
inflation and plays the role of dark energy in the late-time of the universe. We show that our deflationary scenario is applicable 
to arbitrary slow-roll inflation models. We also take two specific inflation potentials to illustrate our results.
\end{abstract}

\maketitle

\section{Introduction} \label{sec:introduction}

It is well-known that there exist two accelerating expansions in our universe.
One occurs at the very early time of the universe, solving particularly the flatness and  horizon problems, 
called ``Inflation''~\cite{Starobinsky:1980te, Guth:1980zm, Starobinsky:1982ee, Linde:1983gd}, and
the other is at the late-time, indicated by the type-Ia supernovae observations~\cite{Riess:1998cb,Perlmutter:1998np}, 
called ``Dark Energy.''
The former epoch can be realized by introducing the inflaton field with a slow-roll potential.
For the latter,
various theoretical  models have been proposed   to achieve the late-time acceleration universe~\cite{Copeland:2006wr},
while the simplest one is to keep the cosmological constant, $\Lambda$, in  the gravitational theory, 
referred to as the $\Lambda$CDM model.
However, if $\Lambda$ is originated from
the vacuum energy, it is associated with the Planck mass  as predicted in particle physics,
which is about $10^{123}$ orders of magnitude larger than the current measured value.
This ``fine-tuning'' or ``hierarchy'' problem in fact was known even before the discovery of dark energy~\cite{WBook}.

There have been many  attempts to solve the hierarchy problem~\cite{Review1}.
One of the popular ways is the running $\Lambda$ model, in which the vacuum energy decays to 
matter~\cite{Ozer:1985ws, Carvalho:1991ut, Lima:1994gi, Lima:1995ea, Overduin:1998zv, Carneiro:2004iz, Shapiro:2004ch, Bauer:2005rpa, Alcaniz:2005dg, Barrow:2006hia, Shapiro:2009dh} 
or a quintessence field~\cite{Costa:2007sq} in the evolution of the universe, and its observational tests have been extensively 
investigated in the literature~\cite{EspanaBonet:2003vk, Borges:2007bh, Borges:2008ii, Tamayo:2015qla, Sola:2016vis}.
In this study, we concentrate on a simple model with the vacuum energy  non-minimally coupled to the inflaton field,
in which the coupling may arise from the conformal transformation, i.e., transforming the Brans-Dicke theory 
from the Jordan frame to the Einstein one~\cite{Peebles:1999fz, Copeland:2000hn, Sahni:2001qp, Sami:2004xk, 
Hossain:2014zma, Geng:2015haa}.
It is interesting to 
  mention that the inflaton coupled to the vacuum energy can be used to realize  inflation 
  with a small coupling constant~\cite{Nishioka:1992sg}.
  
In our model, the vacuum energy begins with the Planck mass scale and deflates by decaying to the inflaton in the pre-inflationary and inflationary epochs.
The non-minimal coupling between the inflaton and vacuum plays the role of heating up the inflaton and triggers inflation, 
whereas  inflation itself is driven by another slow-roll potential.
After the reheating era, the inflaton decays to  matter and decouples to the vacuum energy.
As a result, the residual vacuum energy density is much smaller than the matter one after inflation.
 At the late-time of the universe,
the vacuum energy is dominated again, known as dark energy.
In this scenario, the energy difference between the Planck mass and current cosmological constant scale is determined by the inflaton potential and the coupling constant between the vacuum energy and  inflaton field.
We will  show that the allowed range of the coupling constant is  insensitive to the choice of  the potentials so that
our deflation scenario for $\Lambda$ is quite general.

This paper is organized as follows:
In Sec.~\ref{sec:model}, we introduce the model with the vacuum energy non-minimally coupled to the inflaton field.
In Sec.~\ref{sec:analy_numer}, we calculate the analytical solution and estimate the range of the coupling constant.
In Sec.~\ref{sec:number}, we use two specific potentials to check  our analytical results.
We present our conclusions in Sec.~\ref{sec:conclusion}.

\section{Non-minimally coupled vacuum energy and inflaton}
\label{sec:model}
We start from the Brans-Dicke action with the vacuum energy~\cite{Hossain:2014zma},
\begin{eqnarray}
\label{eq:BD_theory}
S_{\mathrm{BD}}=\int \sqrt{-\tilde{g}} d^4x \left\{ \frac{M_p}{2} \left[ \varphi \tilde{R} - \frac{\omega_{BD}}{\varphi} \tilde{g}^{\mu \nu} \partial_{\mu} \varphi \partial_{\nu} \varphi -2 U(\varphi) \right] + \mathcal{L}_{\Lambda} \right\} \,,
\end{eqnarray}
where $M_p^2=(8 \pi G)^{-1}$ is the Planck mass, $\tilde{g}_{\mu \nu}=(-,+,+,+)$ is the metric in the Jordan frame, 
$\omega_{BD}$ is the Brans-Dicke parameter, and $\mathcal{L}_{\Lambda} \sim - M_{p}^4$ is the Lagrangian density of 
the vacuum energy.
By using the conformal transformation
\begin{eqnarray}
\label{eq:CT-1}
\tilde{g}^{\mu \nu}  = A^{-2} g^{\mu \nu}
\end{eqnarray}
with 
\begin{eqnarray}
\label{eq:CT-2}
\varphi = A^{-2}\,,\quad \left(\frac{1}{A}\frac{dA}{d\phi} \right)^2=\frac{1}{4\varphi}\left( \frac{d\varphi}{d\phi} \right)^2=\frac{4 \pi G}{2 \omega_{BD} +3}\,, \quad V(\phi)= \frac{U(\varphi)}{\varphi^2} \,,
\end{eqnarray} 
the Brans-Dicke action is transformed from the Jordan frame into the Einstein one, given by
\begin{eqnarray}
\label{eq:action_Ein}
S_E= \int \sqrt{-g} d^4x \left[ \frac{M_p^2}{2}R - \frac{\left( \nabla \phi \right)^2}{2} -V(\phi) + \mathcal{L}_{\Lambda}(A^2(\phi) g_{\mu \nu}) \right] \,,
\end{eqnarray}
where $g_{\mu \nu}$ is the Einstein frame metric, $\phi$ is the inflaton field, and $V(\phi)$ is a slow-roll inflaton potential.

By varying the action~(\ref{eq:action_Ein}) with respect to the metric and specializing to the FLRW
case with $g_{\mu \nu} = \mathrm{diag}(-1, a^2, a^2, a^2)$, we obtain,
\begin{eqnarray}
\label{eq:Friedmann-1}
&& H^2= \frac{1}{3M_p^2} \left( \rho_{\phi} + \rho_v \right) \,, \\
\label{eq:Friedmann-2}
&& \dot{H} = - \frac{1}{2M_p^2} \left(  \rho_{\phi}+  P_{\phi} + \rho_v + P_v \right) \,,
\end{eqnarray}
where $\rho_{\phi}$ and $P_{\phi}$ are the energy density and pressure of the inflaton, defined by
\begin{eqnarray}
\label{eq:inflaton_rho_p}
\rho_{\phi} = \frac{\dot{\phi}^2}{2} + V(\phi)\,, \quad P_{\phi} = \frac{\dot{\phi}^2}{2} - V(\phi)\,,
\end{eqnarray}
while $\rho_v$ and $P_v$ are the energy density and pressure of the vacuum, respectively.
Note that the vacuum energy equation of state (EoS) satisfies the relation, $w_v \equiv P_v/\rho_v=-1$.
If we take the conformal transformation coefficient $A^2(\phi)$ in Eq.~(\ref{eq:action_Ein}) to be,
 \begin{eqnarray}
 \label{eq:A2}
A^2(\phi) = e^{ -2\lambda \phi / M_p }\,,
\end{eqnarray}
with $\lambda$ being the model parameter to be determined, the inflaton field and the continuity 
equations for the vacuum energy can be derived as,
\begin{eqnarray}
\label{eq:field_phi-1}
&&\ddot{\phi}+3H\dot{\phi}+\frac{\d V}{\d \phi} = \frac{\lambda}{M_p} (\rho_v-3P_v) \,, \\
\label{eq:cont_v}
&&\dot{\rho}_v+3H(\rho_v+P_v)= - \frac{\lambda}{M_p} \dot{\phi}
(\rho_v-3P_v) \,,
\end{eqnarray}
respectively.
Thus, the vacuum energy can be solved as the function of $\phi$ from Eq.~(\ref{eq:cont_v}),
\begin{eqnarray}
\label{eq:rho_v}
 \rho_v(\phi) = \rho_{\Lambda_o} e^{-4\lambda \phi / M_p} \,,
\end{eqnarray}
where $\rho_{\Lambda_o} $ is the vacuum energy density after the Big Bang.
Consequently, the inflaton field equation in Eq.~(\ref{eq:field_phi-1}) becomes
\begin{eqnarray}
\label{eq:field_phi-2}
\ddot{\phi}+3H\dot{\phi}+\frac{\d V_{\mathrm{tot}}}{\d \phi} = 0 \,,
\end{eqnarray}
where the combined potential is given by
\begin{eqnarray}
\label{eq:effective_V}
V_{\mathrm{tot}}(\phi)=V(\phi)+V_{\mathrm{eff}}(\phi)
\end{eqnarray}
with $V_{\mathrm{eff}}(\phi)= \rho_{\Lambda_o} e^{-4\lambda \phi / M_p}$.

\section{Analytical estimations with generic potential}
\label{sec:analy_numer}

In Eq.~(\ref{eq:A2}), we consider $\lambda>0$, so that the vacuum energy decays to the inflaton field in the pre-inflation and inflation eras.
After the reheating epoch, the inflaton decays to standard model particles, and the vacuum energy decouples to the inflaton field.
As result, the original cosmological constant $\Lambda_o$ at the Planck mass scale is deflated to the current small measured value of $\Lambda_{de}$.
Explicitly, we have
\begin{eqnarray}
\label{eq:phi_min}
 \rho_v(\phi_m)=\rho_{\Lambda_o} e^{-4\lambda \phi_m / M_p} = \rho_{\Lambda_{de}}  \,,
\end{eqnarray}
where $\phi_m$ is the minimum of the inflaton potential $V(\phi)$ and 
 $\rho_{\Lambda_{de}} \simeq 10^{-47}~\mathrm{GeV}^4 \sim 10^{-123} \rho_{\Lambda_o}$ is the current 
dark energy density.
 From Eq.~(\ref{eq:phi_min}), we get
\begin{eqnarray}
\label{eq:phi_min1}
\frac{\phi_m}{M_p} \simeq \frac{71}{\lambda} \,.
\end{eqnarray}

\begin{figure}
\centering
\includegraphics[width=0.49 \linewidth]{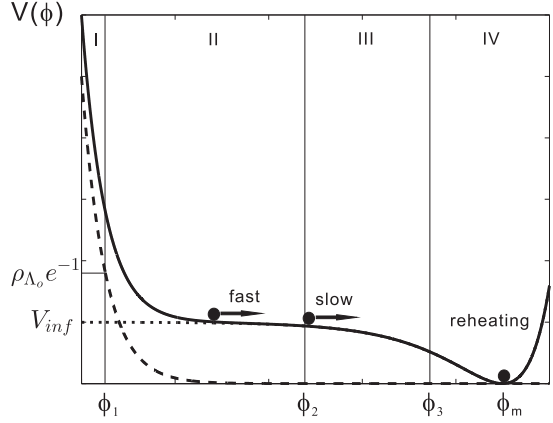}
\caption{The potential $V(\phi)$ versus the inflaton $\phi$, where the solid line denotes the effective potential $V_{\mathrm{eff}}$ in Eq.~(\ref{eq:effective_V}), while the dashed and dotted lines represent  the  terms from the coupled vacuum energy and  slow-roll potential, respectively.}
\label{fg:1}
\end{figure}

In Fig.~\ref{fg:1}, we illustrate the potential $V(\phi)$ as a function of the inflaton $\phi$, where the solid, dashed and dotted lines correspond to the combined potential $V_{\mathrm{tot}}$, the effective potential $V_{\mathrm{eff}}$ and the slow-roll potential $V(\phi)$ in Eq.~(\ref{eq:effective_V}), respectively.
The evolution history can be divided into the following four stages.

\vspace{0.5em}%
\noindent {\em (i) the epoch of the vacuum energy decay}

At the first stage, the vacuum energy decays to the inflaton field 
with $\rho_\phi+\rho_v\simeq constant$
and ends up at $\phi=\phi_1$ with $\rho_v = \rho_{\Lambda_o} e^{-1}$ (see also Fig.~\ref{fg:1}).
In this period, the evolution of $\phi$ is dominated by $V_{eff}(\phi)$, indicating that the inflaton is heating up by the vacuum energy decay.
The growth of the inflaton $\phi_1$ is given by
\begin{eqnarray}
\label{eq:phi1}
\rho_v= \rho_{\Lambda_o} e^{-4\lambda \phi_1 /M_p} = \rho_{\Lambda_o} e^{-1} \quad \Rightarrow \quad \frac{\phi_1}{M_p}= \frac{1}{4 \lambda} \,,
\end{eqnarray}
which is much smaller than $\phi_m$ in Eq.~(\ref{eq:phi_min1}).
We note that since $\rho_\phi>\rho_v$ and $\rho_\phi\rightarrow O(M_p^4)$
by the end of this stage, the energy flow from the vacuum energy to the inflaton 
slightly influences the evolution of $\rho_\phi$ after this stage.

\vspace{0.5em}%
\noindent {\em (ii) the fast-roll pre-inflationary epoch.}

In this stage, the universe is undergoing the decelerating pre-inflationary era and dominated by the ``hot'' inflaton, corresponding to $\dot{\phi}^2 \gg V(\phi)$.
The energy density follows the continuity equation and is diluted by the expansion of the universe with $\rho_{\phi} \propto a^{-3(1+w_{\phi})}$.
This stage is terminated when the slow-roll inflation occurs at $\phi = \phi_2$, i.e., $\dot{\phi}^2_2 \ll V(\phi_2)$, leading to 
\begin{eqnarray}
\rho_{\phi}(\phi_2) \simeq V(\phi) \equiv V_{inf} \,,
\end{eqnarray}
where $V_{inf}$ is the potential energy during inflation (see also Fig.~\ref{fg:1}).
If we consider that the value of $V(\phi)$ at the beginning and end of inflation are of the same order of magnitude, the scale of the potential energy is $V_{inf} \simeq \rho_{\phi =\phi_{inf}} \simeq M_p^2 H_{inf}^2 \simeq M_p^2 H_{end}^2$, where $H_{inf(end)}$ is the Hubble parameter at the inflationary-era (end) of inflation with $H_{end} \sim 10^{-6} M_p$ given in Ref.~\cite{Liddle:1993ch}.

By substituting Eqs.~(\ref{eq:Friedmann-1}) - (\ref{eq:inflaton_rho_p}) into Eq.~(\ref{eq:field_phi-2}) with $\dot{\phi}^2 \gg V(\phi)$, we
have
\begin{eqnarray}
\label{eq:dphi_2}
\frac{\phi^{\prime}}{M_p} \equiv \frac{1}{M_p} \frac{d \phi}{d \ln a} = \sqrt{6} \,.
\end{eqnarray}
Note that the equation of state of the inflaton is given by
\begin{eqnarray}
w_{\phi} = \frac{P_\phi}{\rho_\phi} = \frac{\frac{\dot{\phi}^2}{2}-V}{\frac{\dot{\phi}^2}{2}+V} \simeq 1
\end{eqnarray}
in this stage.
The inflaton energy decreases from the Planck scale to the inflation scale at the end of (ii), resulting in that
\begin{eqnarray}
\frac{\rho_{\phi}(\phi_2)}{\rho_{\Lambda_o}} = \left(\frac{a_2}{a_1}\right)^{-3(1+w_\phi)} \simeq \frac{V(\phi_2)}{\rho_{\Lambda_o}} \simeq 10^{-12} \,,
\end{eqnarray}
which gives the e-folding during this stage to be $N_2\equiv \ln (a_2/a_1) \simeq 4.6$.
Combining $N_2$ with Eq.~(\ref{eq:dphi_2}), we find that the stage (ii) is terminated at $\Delta \phi_2 /M_p = (\phi_2-\phi_1)/ M_p \simeq N_2 \times \phi^{\prime} / M_p \simeq 11.3 $.
We note that the inflationary energy scale is model-dependent, but the choice is insensitive to $\phi_2$.
For example, if $H_{inf} = 10^2 H_{end} = 10^{16}$~GeV and $\Delta \phi_2 = 7.5$, the growths of $\phi_{1,2}$ are of the same order of magnitude.

\vspace{0.5em}%
\noindent {\em (iii) the slow-roll inflationary epoch.}

In this epoch,
inflation is triggered and the evolution of the universe depends on the inflation model.
This stage ends up with the increasing of the e-folding $N_3 \simeq 60$ and
\begin{eqnarray}
\epsilon \lvert_{\phi_3} \equiv -\frac{\dot{H}}{H^2} \lvert_{\phi_3} = 1 \,,
\end{eqnarray}
where $\epsilon$ is the slow-roll parameter.
The growth of the inflaton $\Delta \phi_3 = \phi_3 - \phi_2$ is extremely model-dependent in this period.
However, it is still possible to estimate its value by taking the slow-roll condition, i.e.
\begin{eqnarray}
\ddot{\phi} &+& 3H\dot{\phi} + \frac{dV}{d\phi} \simeq 3H\dot{\phi} + \frac{dV}{d\phi} \simeq 0\,, \nonumber \\
&\Rightarrow & \frac{\phi^{\prime}}{M_p} \simeq - M_p \frac{V_\phi}{3H^2} \simeq - \frac{1}{M_p} \frac{V_\phi}{V} = \sqrt{2 \epsilon} \,,
\label{eq:dphi_3}
\end{eqnarray}
 with
\begin{eqnarray}
\label{eq:epsilon}
\epsilon = -\frac{\dot{H}}{H^2} \simeq \frac{M_p^2}{2} \left( \frac{dV/d\phi}{V} \right)^2 \,.
\end{eqnarray}
In addition, the tensor-to-scalar ratio is calculated by $r=16 \epsilon$.
Clearly, it is reasonable for us to have the estimation of
\begin{eqnarray}
\frac{\phi^{\prime}}{M_p} \simeq \sqrt{\frac{r}{8}} \lesssim 0.15
\end{eqnarray}
with $r<0.18$ in  this stage.
As a result, the growth of the inflaton is $\Delta \phi_3 / M_p \lesssim 9$ by taking $N=60$ during  inflation.
We note that the allowed $\Delta \phi_3$, depending on the specific inflation model, can be bigger, but the order
of magnitude  should be the same.
By considering various reasonable parameters,
we take
\begin{eqnarray}
{\Delta \phi_3\over M_p} \lesssim 18 \,.
\end{eqnarray}

\vspace{0.5em}%
\noindent {\em (iv) the reheating epoch}

Here, $\Delta \phi_4=(\phi_m-\phi_3)$ is also model dependent.
We assume that the potential can be expressed as $V \lvert_{\phi \rightarrow \phi_m} \sim \phi^2$ around the potential minimum.
If we take $\epsilon \lvert_{\phi_3} = 1$ as the condition at the end of inflation, we can deduce that $\Delta \phi_4/M_p=(\phi_m-\phi_3)/M_p \sim \sqrt{2}$.

As a result, $\phi_m$ is added up by the growth of $\phi$ from (i) to (iv), given by
\begin{eqnarray}
\label{eq:phi_min2}
12.7 \lesssim \phi_m /M_p = (\phi_1 + \Delta \phi_2 + \Delta \phi_3 + \Delta \phi_4)/M_p \lesssim 30.7 \,.
\end{eqnarray}
From Eqs.~(\ref{eq:phi_min1}) and (\ref{eq:phi_min2}), we can roughly estimate the allowed range 
of the coupling constant $\lambda$ to be
\begin{eqnarray}
2.3 \lesssim \lambda \simeq \frac{71 M_p}{\phi_m} \lesssim 5.6 \,.
\end{eqnarray}

\section{Numerical results with specific potentials}
\label{sec:number}

To check our estimations in Sec.~\ref{sec:analy_numer}, we present the numerical evaluations with two specific  slow-roll 
potentials:
\begin{eqnarray}
\label{potential1}
V(\phi) &=& V_0 \left(1-e^{\sqrt{\frac{2}{3}}(\phi-\phi_m)/M_p}\right)^2\,,
\\
\label{potential2}
V(\phi) &=& m^2 \left(\phi-\phi_m\right)^2\,.
\end{eqnarray}
Note that the first potential in Eq.~(\ref{potential1}) can be obtained from 
 the Starobinsky's $R^2$ inflation model~\cite{Starobinsky:1980te}
 after the conformal transformation 
 from the Jordan to  Einstein frame.

\begin{figure}
\centering
\includegraphics[width=0.49 \linewidth]{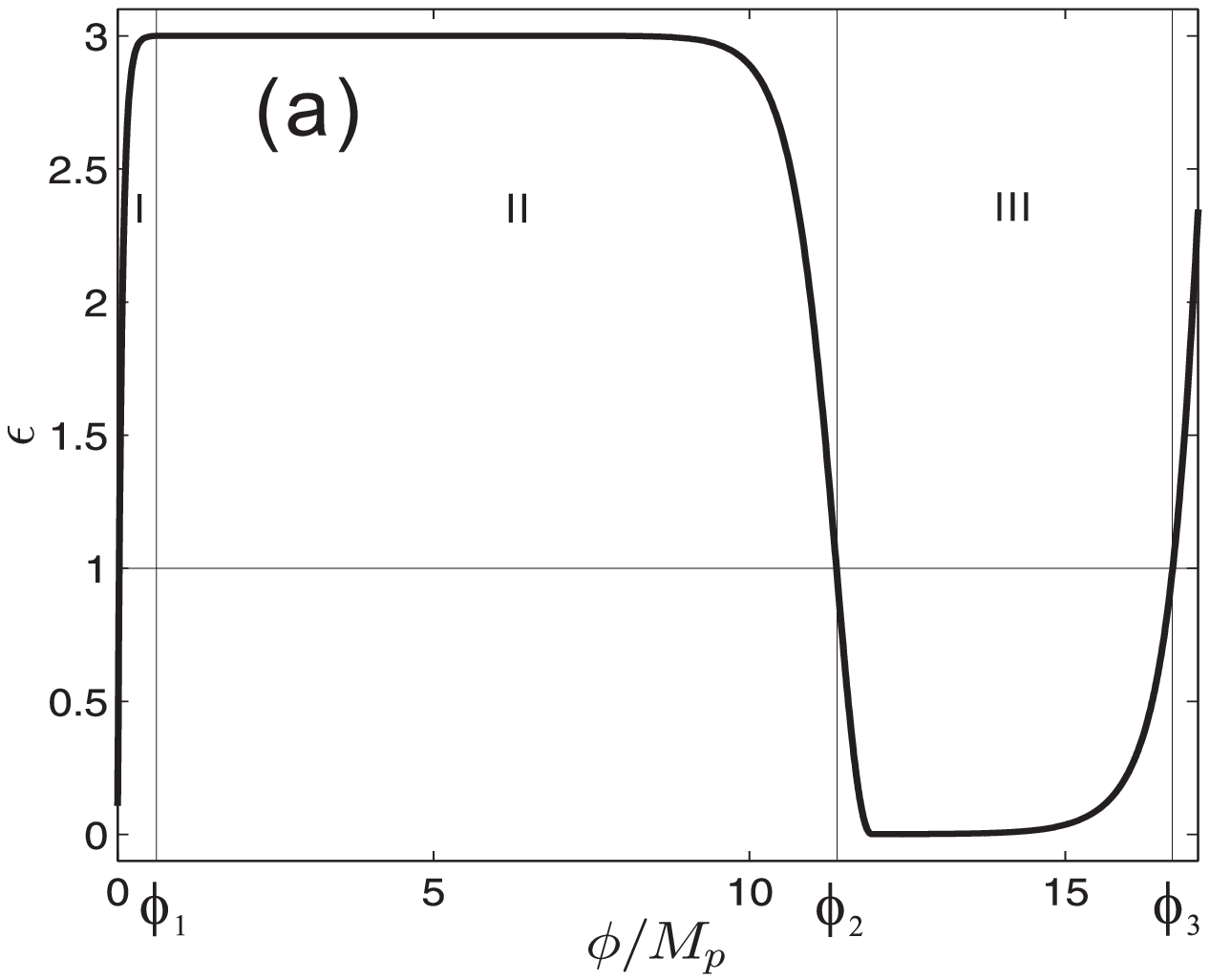}
\includegraphics[width=0.49 \linewidth]{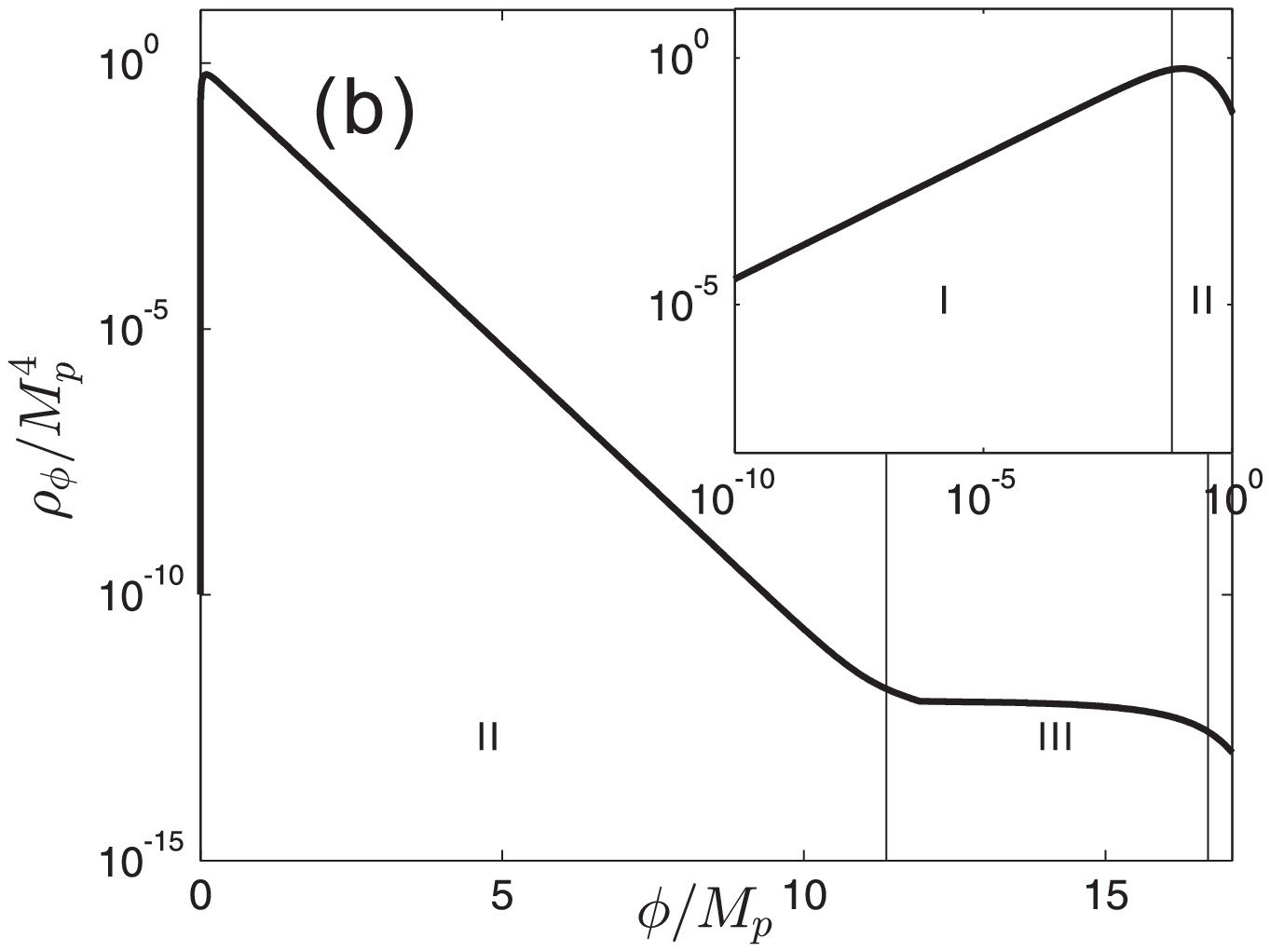}
\caption{ 
Evolutions of (a) $\epsilon$ and (b) $\rho_{\phi} / M_p^4$ as functions of $\phi$ with $V(\phi) = V_0 (1-e^{\sqrt{\frac{2}{3}} (\phi-\phi_m)/M_p})^2$, $V_0=10^{16}~\mathrm{GeV}^4$ and $\lambda=4.09$.
}
\label{fg:2}
\end{figure}
\begin{figure}
\centering
\includegraphics[width=0.49 \linewidth]{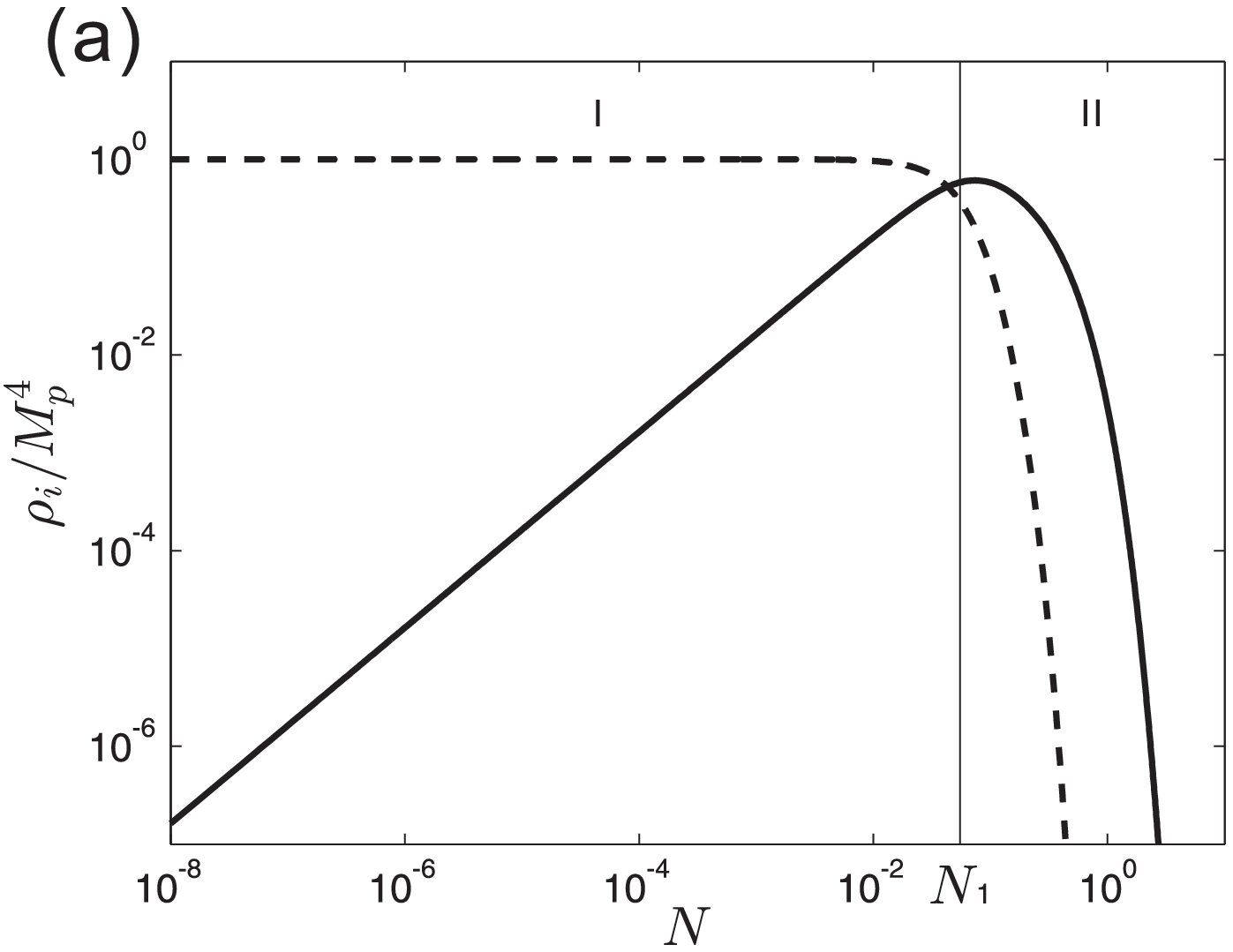}
\includegraphics[width=0.49 \linewidth]{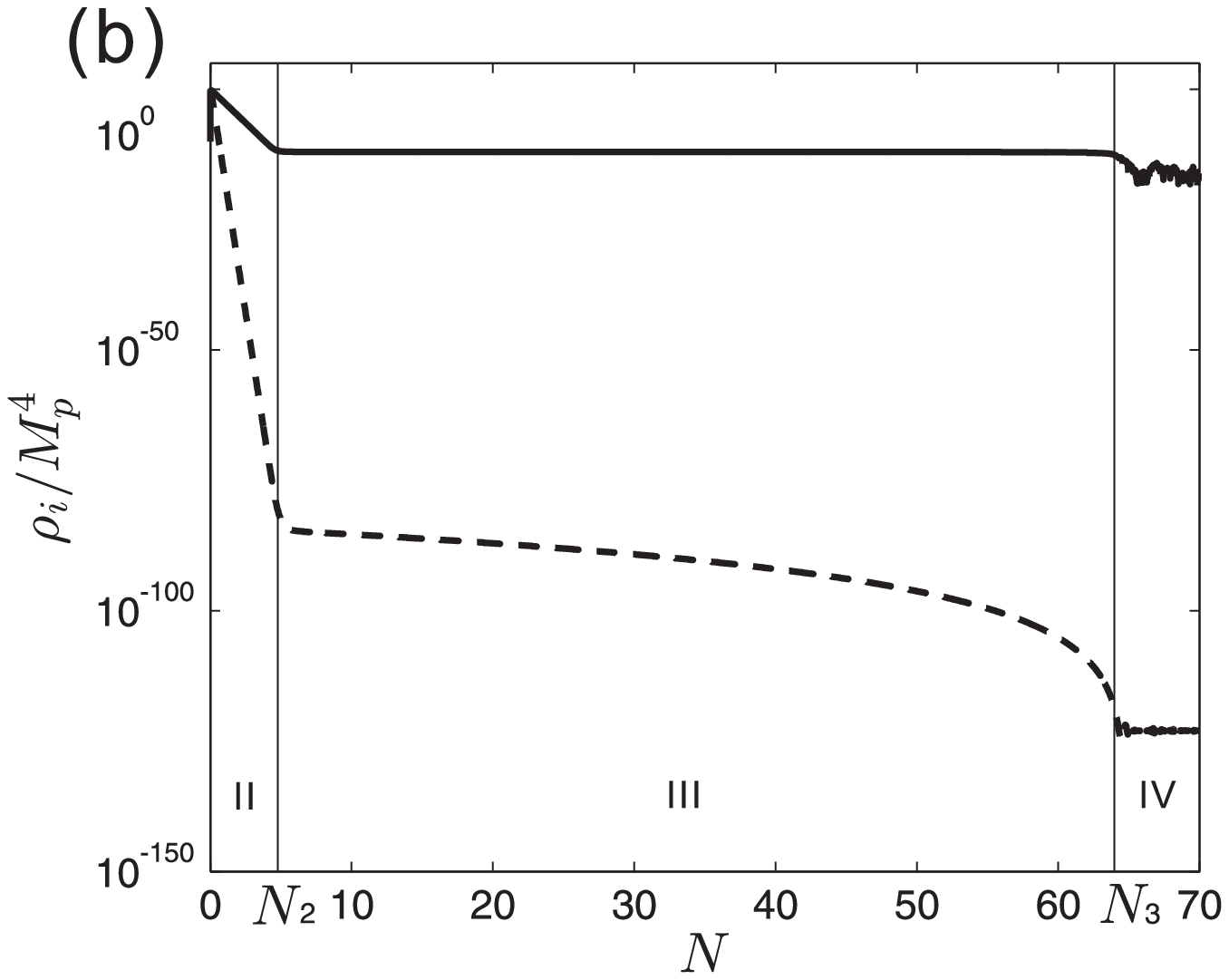}
\caption{ 
Evolutions of $\rho_i / M_p^4$ with $i=\phi$ (solid) and $v$ (dashed) as functions of the e-folding $N$ 
for  (a) $10^{-8}<N<10$ and (b) $N<70$.
Legend is the same as Fig.~\ref{fg:2}.
}
\label{fg:5}
\end{figure}
In Fig.~\ref{fg:2}, we demonstrate the evolutions of
(a) the slow-roll parameter $\epsilon \equiv - \dot{H}/H^2$ and (b) $\rho_{\phi}$ as functions of $\phi$ with the potential in Eq.~(\ref{potential1}) and $\lambda=4.09$.
From the figure, we can see the evolution behaviors of $\phi$ from the stage (i) to (iv).
Note that we have taken that inflation occurs at $\epsilon \lvert_{\phi_2} =1$ and ends up at $\epsilon \lvert_{\phi_3} =1$, while the mass hierarchies are 
 $V_{inf}/M_p^4 = 10^{-12}$ and $\rho_{\Lambda_{de}} / \rho_v^{(0)} = 10^{-123}$.
In Fig.~\ref{fg:5}, we illustrate the evolutions of $\rho_v$ and $\rho_{\phi}$ as functions of the e-folding $N$.
From the plots, we can see that the decaying vacuum energy heats up the inflaton in the stage (i) with $N_1 = 0.06$.
The inflaton energy approaches the Planck scale at the beginning of the stage (ii) and is diluted by the expansion of the universe as $\rho_{\phi} \sim a^{-6}$ with $N_2=4.78$ and $\rho_{\phi}(\phi_2)/M_P^4=1.7 \times 10^{-12}$.
In addition, we observe that $\rho_{\phi}$ keeps to be constant in the inflationary era in the stage (iii) until $N_3=63.98$ and evolves to the reheating epoch in the stage (iv).
On the other hand, $\rho_v$ decreases in the stages (i - iii), whereas it is nearly constant in the stage (iv), implying that the detail of the reheating is insensitive to the final value of $\rho_v$.

\begin{figure}
\centering
\includegraphics[width=0.49 \linewidth]{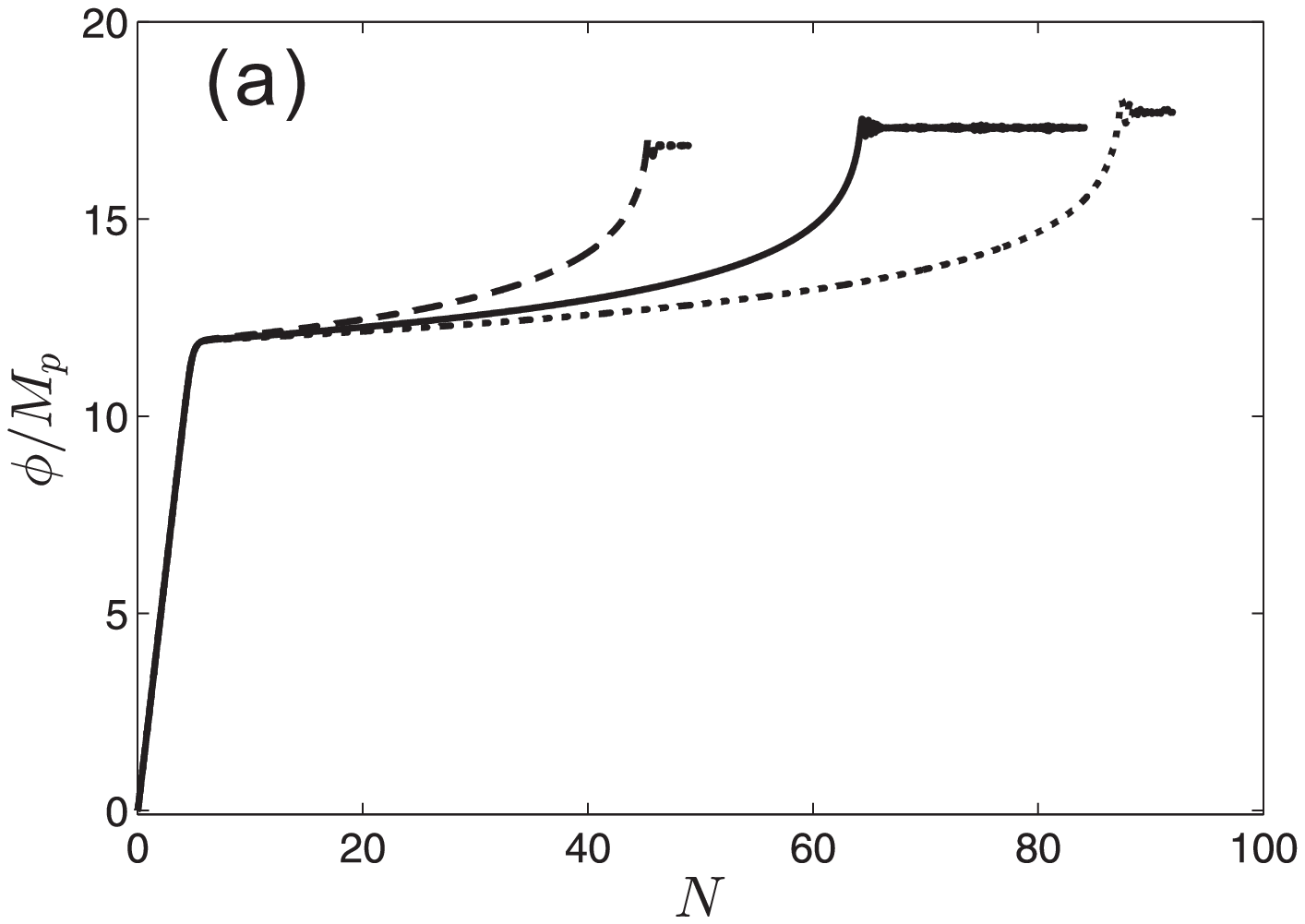}
\includegraphics[width=0.49 \linewidth]{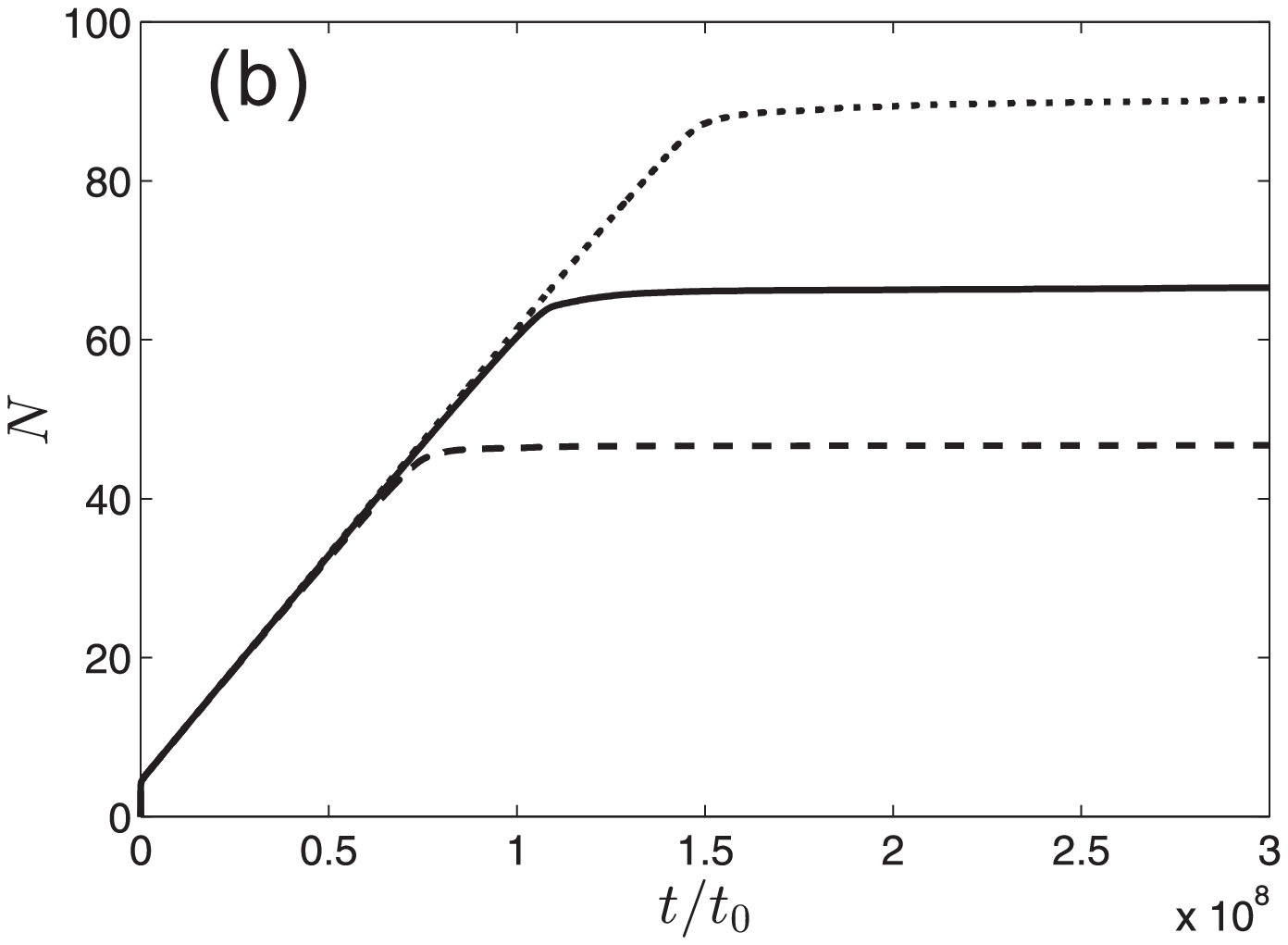}
\caption{
Evolutions of (a) the inflaton $\phi$ as a function of the e-folding $N$ and (b) $N$ as a function of $t/t_0$ with $V(\phi) = V_0 (1-e^{\sqrt{\frac{2}{3}} (\phi-\phi_m)/M_p})^2$ and $\lambda=$ 4.09 (solid), 4.20 (dashed) and 4.00 (dotted), where $t_0=M_p^{-1}$.
}
\label{fg:3}
\end{figure}
In Fig.~\ref{fg:3}, we use $\lambda=$ (4.09, 4.20, 4.00), which are represented by
the solid, dashed and dotted lines, respectively, and $V_0/M_p^4=10^{-12}$, corresponding to the inflation energy at $10^{12}~$GeV.
In Fig.~\ref{fg:3}a, we plot the $\phi$ evolution in terms of the e-folding $N= \ln a$, and we can observe that 
inflation occurs at $\phi_2 \simeq 11.4$ and ends up at $\phi_3 \simeq 16.7$, which fit our predictions of 
 $\phi_2=11.3$ and $\Delta \phi_3 \lesssim 18$.
Fig.~\ref{fg:3}b shows the e-folding $N$ versus the normalized physical time $t/t_0$, where $t_0=M_p^{-1}$ is the Planck time, and the corresponding e-foldings are $N=59.2$, $40.1$ and $82.3$, respectively.
The figure indicates that the e-folding $N$ is one-to-one correspondence to the coupling constant $\lambda$.

\begin{figure}
\centering
\includegraphics[width=0.49 \linewidth]{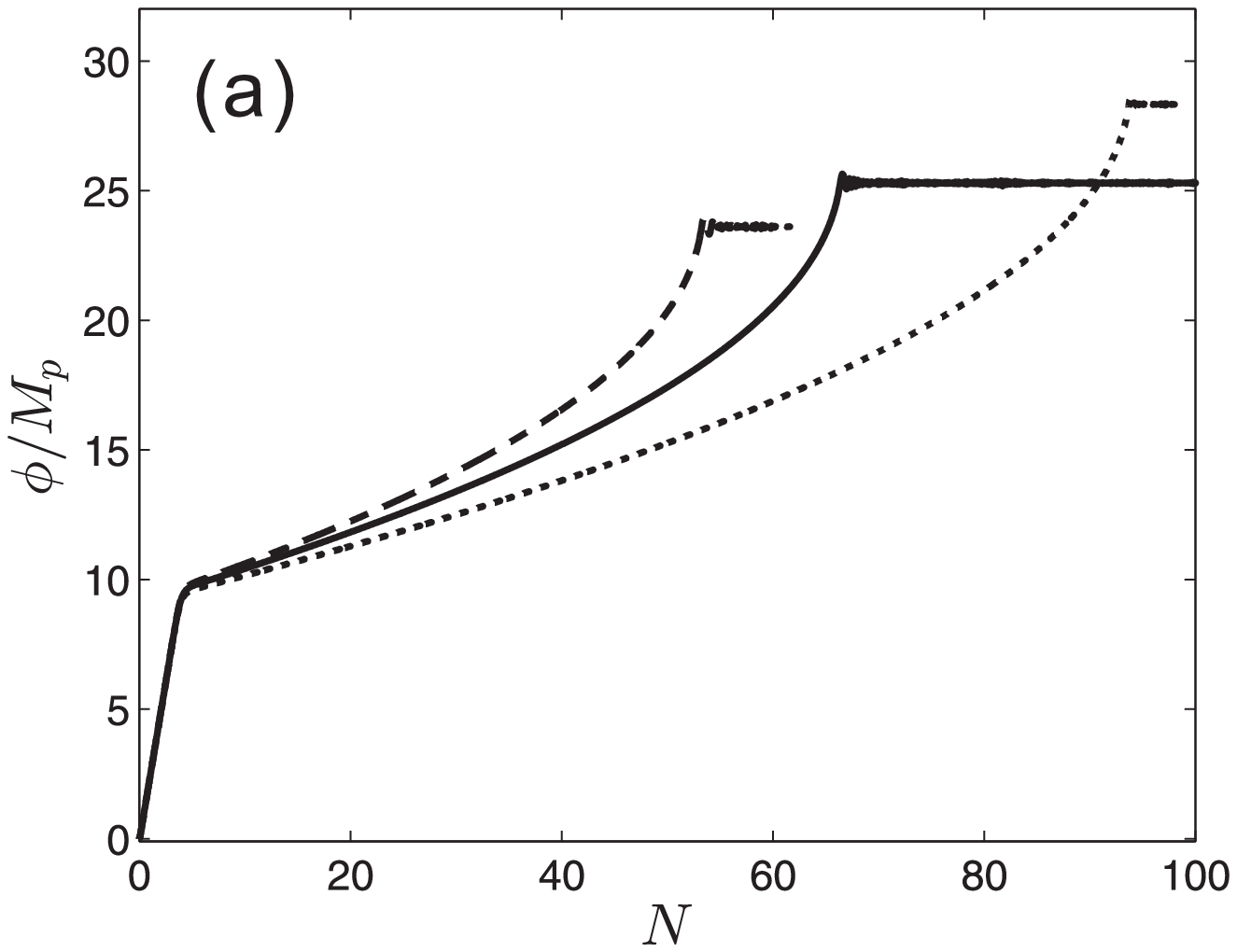}
\includegraphics[width=0.49 \linewidth]{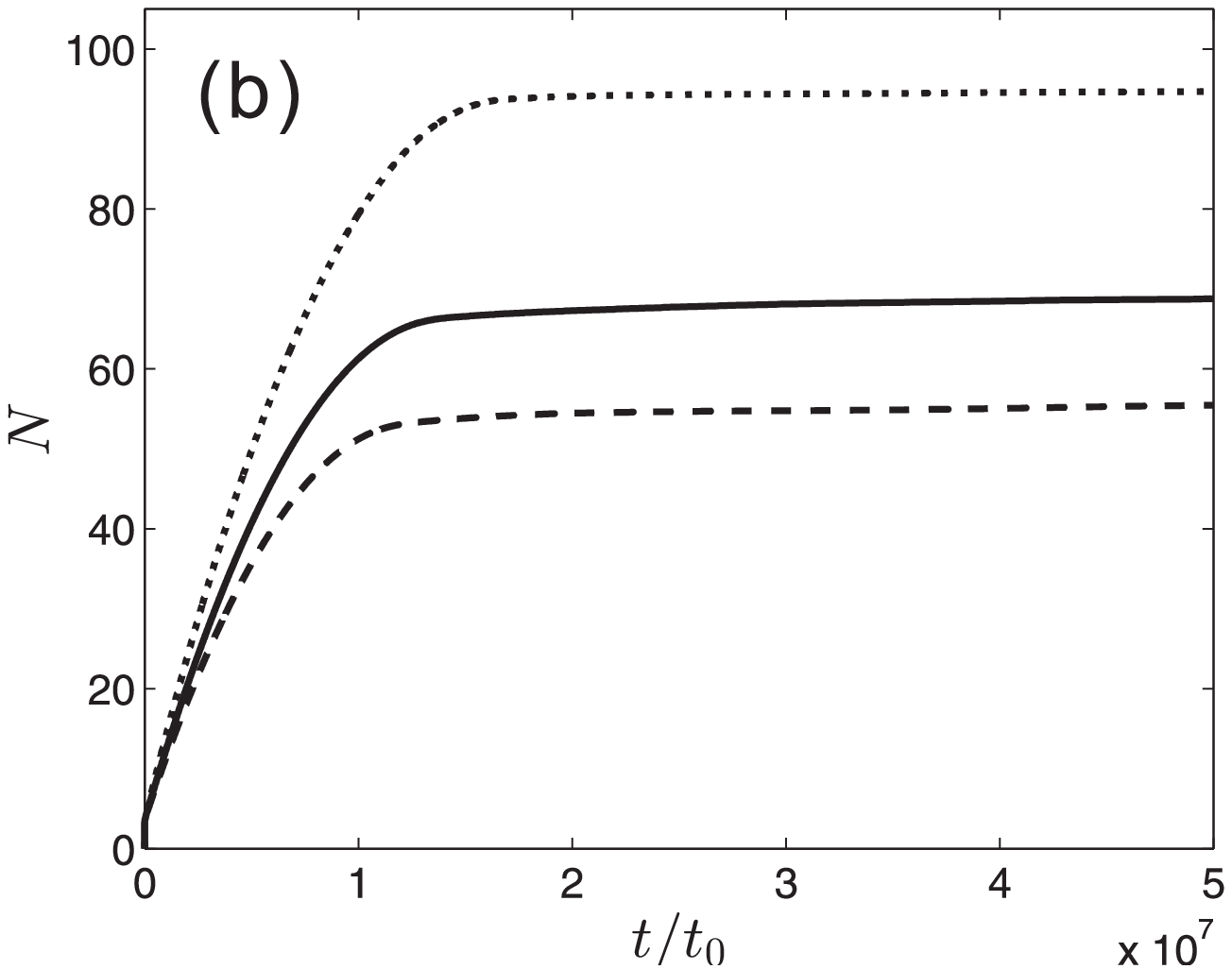}
\caption{
Legend is the same as Fig.~\ref{fg:3} but
$V(\phi) = m^2 (\phi-\phi_m)^2$ and $\lambda=$ 2.5 (solid), 2.75 (dashed) and 2.25 (dotted).
}
\label{fg:4}
\end{figure}

We now examine the large field inflation potential $V(\phi)=m^2 \phi^2$~\cite{Linde:1983gd} in Eq.~(\ref{potential2})
with $m=10^{13}~$GeV. 
In Fig.~\ref{fg:4}, we plot the evolutions of (a) $\phi$ and (b) $N= \ln a$ as functions of $N$ and $t/t_0$, respectively.
In the figure, we choose $\lambda=(2.8, 3.0, 2.5)$, denoted by solid, dashed and  dotted lines, respectively.
From Fig.~\ref{fg:4}a, we find that inflation happens and ends up at $(\phi_2,~\phi_3) / M_p \simeq (9.1,~24.3)$ with $N = 60$,
resulting in $\Delta \phi_3 \simeq 15.2$, which is the same order as our expectation.
As shown in Fig.~\ref{fg:4}b,  the corresponding e-foldings are $N=61.9$, $48.9$ and $89.4$, respectively,
which also illustrate the one-to-one correspondence to the coupling constant $\lambda$.

\section{Conclusions}
\label{sec:conclusion}

We have proposed  a deflationary cosmological constant model to understand the hierarchy problem, in which
the   vacuum energy non-minimally couples to the inflaton field.
In this model, the energy difference between the Planck mass and the current scale of the cosmological constant 
is determined by the non-minimal coupling constant $\lambda$ and the inflaton at the minimum of the potential $\phi_m$.
These two parameters are no longer to be unreasonable huge (or small), so that the hierarchy problem can be resolved.
Explicitly, 
we have estimated that the allowed range of $\lambda$ and $\phi_m$ are around  $2.3 \lesssim \lambda \lesssim 5.6$ and $12.7 \lesssim \phi_m / M_p \lesssim 30.7$, which depend mildly on the  inflation models.
Our results have also been supported by the numerical calculations of the two popular slow-roll potentials.

\section*{Acknowledgments}
This work was partially supported by National Center for Theoretical Sciences,  National Science Council (NSC-101-2112-M-007-006-MY3), MoST (MoST-104-2112-M-007-003-MY3)  and National Tsing Hua University~(104N2724E1).

\end{document}